\UseRawInputEncoding
\documentclass[aps,pra,floatfix,superscriptaddress,twocolumn,showpacs,10pt]{revtex4-1}
\usepackage{amssymb, amsmath,color,graphicx,subfigure}
\usepackage{calc,epsfig,epstopdf,bm,mathrsfs}
\usepackage{times}
\usepackage{multirow}

\usepackage{CJKutf8}
\begin{document}
\begin{CJK}{UTF8}{gbsn}

\title{Nonadiabatic Geometric Quantum Gates with on-Demand Trajectories}

\author{Yan Liang}

\affiliation{School of Physical Science and Technology, Guangxi Normal University, Guilin 541004, China}

\affiliation{Key Laboratory of Atomic and Subatomic Structure and Quantum Control (Ministry of Education),  Guangdong Basic Research Center of Excellence for Structure and Fundamental Interactions of Matter, and School of Physics, South China Normal University, Guangzhou 510006, China}

\author{Zheng-Yuan Xue} \email{zyxue83@163.com}
\affiliation{Key Laboratory of Atomic and Subatomic Structure and Quantum Control (Ministry of Education),  Guangdong Basic Research Center of Excellence for Structure and Fundamental Interactions of Matter, and School of Physics, South China Normal University, Guangzhou 510006, China}

\affiliation{Guangdong Provincial Key Laboratory of Quantum Engineering and Quantum Materials,\\    Guangdong-Hong Kong Joint Laboratory of Quantum Matter,  and Frontier Research Institute for Physics,\\ South China Normal University, Guangzhou 510006, China}

\affiliation{Hefei National Laboratory,  Hefei 230088, China}

\date{\today}

\begin{abstract}

High-fidelity quantum gates are an essential prerequisite for large-scale quantum computation. When
manipulating practical quantum systems, environmentally and operationally induced errors are inevitable,
and thus, in addition to being fast, it is preferable that operations should be intrinsically robust against
different errors. Here, we propose a general protocol for constructing geometric quantum gates with on-demand trajectories by modulating the applied pulse shapes that define the system’s evolution trajectory.
Our scheme adopts reverse engineering of the target Hamiltonian using smooth pulses, which also eliminates the difficulty of calculating geometric phases for an arbitrary trajectory. Furthermore, because a
particular geometric gate can be induced by various different trajectories, we can further optimize the gate
performance under different scenarios; the results of numerical simulations indicate that this optimization
can greatly enhance the quality of the gate. In addition, we present an implementation of our proposal
using superconducting circuits, showcasing substantial enhancements in gate performance compared with
conventional schemes. Our protocol thus presents a promising approach for high-fidelity and strong-robust
geometric quantum gates for large-scale quantum computation.

%%%%%%%%%%%%%%%%%%%%%%%%%%%%%%%%%%%%%%%%%%%%%%%%%%%%
\end{abstract}
\maketitle

\section{Introduction}

Quantum computation is an emerging technology that can succeed in dealing with problems that are hard for
classical computers \cite{Neilsen2000}. During recent decades, there has been tremendous progress toward practical quantum computation on various physical platforms \cite{progress, ion, photon, semiconductor, superconductor}. A recent milestone in this field is that the quantum computational advantage of quantum computers has been successfully demonstrated \cite{advantage, advantage2, advantage3, advantage4}; however, it is still difficult to build a practical quantum computer. This is due to the high overhead of quantum error correction, as gates errors can accumulate and propagate \cite{error}, and this will quickly cause the quantum computation process to fail. Current efforts are devoted to successfully extending the lifetimes of logical qubits beyond those of their corresponding physical qubits \cite{logical1, logical2}, and it has been demonstrated that quantum error correction codes can benefit quantum information storage and processing. Furthermore, it is well known that fewer resources are needed for quantum error correction when the quantum gates used have higher fidelity; therefore, high-fidelity quantum gates are essential for building a practical quantum computer.

It is also well known that geometric phases are intrinsically robust against operational errors \cite{robust, robust2, robust3}, and quantum gates induced by geometric phases thus have built-in fault tolerance. In addition, to decrease the time that quantum systems are exposed to the environment, fast nonadiabatic geometric quantum gates (NGQGs) are preferable \cite{nonad, nonad2, nonad3}. During the last two decades, we have seen great progress in this field \cite{loop2,loop3,loop,loop4,loop5,loop6,xu2014protecting,xu2014universal,TChen2020,CYDing20212,liangyan2023,Cheny2023,DLeibfried2003,du2006,YXu2020,PZZhao2021}; however, due to the difficulty of calculating geometric phases for arbitrary trajectories, only specific loops \cite{loop, loop2, loop3, loop4, loop5, loop6, xu2014protecting, xu2014universal}, usually constructed from longitude and latitude lines on the Bloch sphere, are proposed for implementing NGQGs, and these result in sudden changes of the Hamiltonian parameters. These settings introduce more experimental requirements for implementing geometric quantum gates, and more unacceptably, they will smear the intrinsic robustness of geometric phases.

Here, we address the above obstacles by presenting a general protocol to construct NGQGs with on-demand trajectories by modulating the applied pulse shapes. Our scheme adopts a reverse engineering \cite{reverse,Odelin2019,li2020app,zhao2020gener} of the target Hamiltonian using only smooth pulses, avoiding sudden changes in the system’s Hamiltonian, which also removes  the difficulty of calculating geometric phases. We can also further optimize the gate performance among different trajectories under various scenarios, and numerical simulations verify the enhancement in the gate quality. In addition, we present an implementation of our 74
scheme with superconducting circuits and show that it has improved gate performance. Our protocol therefore presents a promising approach for high-fidelity and strong-robust NGQGs for large-scale quantum computation.

 \begin{figure*}[tbp]
  \centering
  \includegraphics[width=\linewidth]{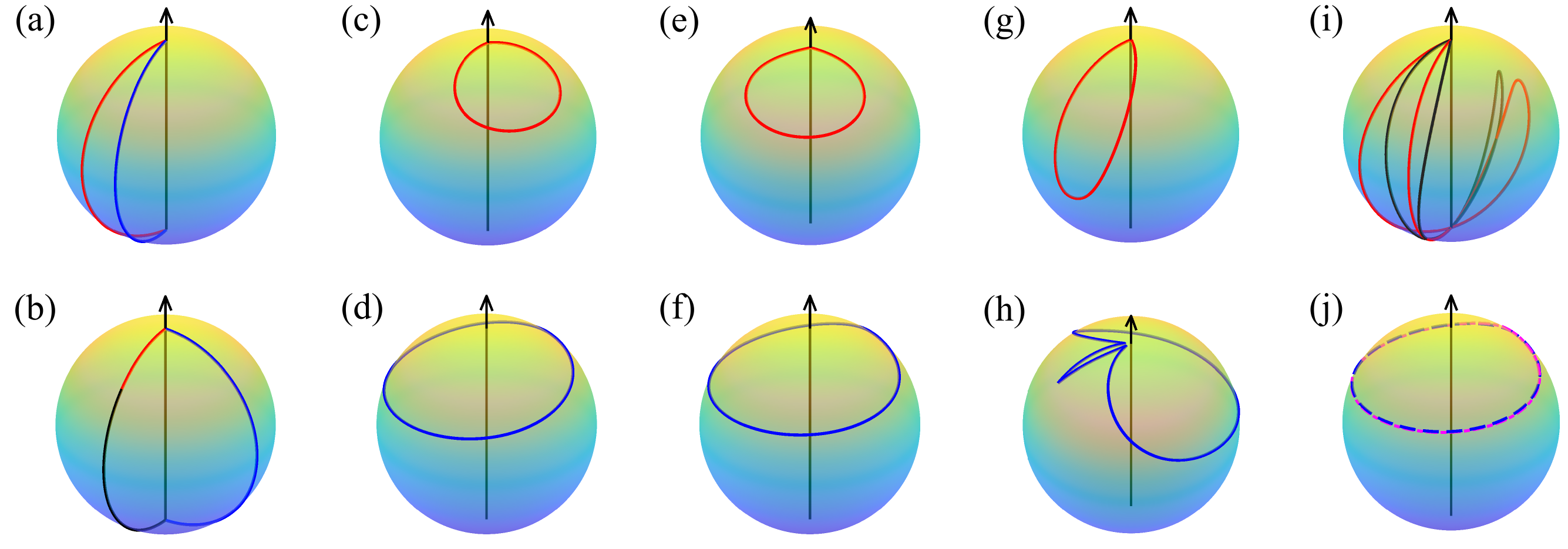}
  \caption{ Evolution paths of $T$ and $H$ gates, with the upper row depicting the T gate and the lower row showcasing the $H$ gate. (a), (b)
Evolution paths of $T$ and $H$ gates of the conventional NGQG scheme. Panels (c), (d), (e), (f), and (g), (h) respectively illustrate the
evolution paths for the $T$ and $H$ gates corresponding to Case 1, Case 2, and Case 3 in the ONGQG scheme. (i) Evolution paths of $T$
gate, with the red line representing Case 4 and the black line representing Case 5. (j) Evolution paths of $H$ gate, with the blue line
representing Case 4 and the purple line representing Case 5.
  }
  \label{path} 
\end{figure*}

\section{Gate construction}

We first proceed to present our scheme for constructing an NGQG using the reverse engineering of the target Hamiltonian. To induce target gates for a quantum two-level system, we choose a set of evolution states $|\psi_{k}(t)\rangle=e^{{\rm i}\gamma_{k}(t)}|\mu_{k}(t)\rangle$  ($k=1,2$) that satisfy the Schr\"{o}dinger equation ${\rm i}|\dot{\psi}_{k}(t)\rangle=H(t)|\psi_{k}(t)\rangle$, where we set $\hbar =1$ hereafter,  $\gamma_{k}(0)=0$ and $\gamma_{k}(t)$ are the accumulated total phase, and
 \begin{eqnarray}
\label{state}
&|\mu_{1}(t)\rangle=\cos\frac{\theta(t)}{2}|0\rangle +\sin\frac{\theta(t)}{2}e^{{\rm i}\varphi(t)}|1\rangle , \notag\\
&|\mu_{2}(t)\rangle=\sin\frac{\theta(t)}{2}e^{-{\rm i}\varphi(t)}|0\rangle-\cos\frac{\theta(t)}{2}|1\rangle,
\end{eqnarray}
are a set of orthogonal auxiliary basis vectors with $\theta(t)$ and $\varphi(t)$ being the time-dependent parameters. We further assume that the quantum system is governed by the following general form of Hamiltonian \cite{li2020app}
\begin{eqnarray}
\label{hamiltonian}
H(t)&=&\!\!{\rm i}\!\sum_{k\neq l \in \{1, 2\}} \langle\mu_{l}(t)|\dot{\mu}_{k}(t)\rangle|\mu_l(t)\rangle\langle\mu_{k}(t)| \notag \\
    &=&\!\Delta(t)  (|0\rangle\langle0|-|1\rangle\langle1|)
      +[\Omega(t)|1\rangle\langle0|+\rm{H.c.}],
\end{eqnarray}
where $\Delta(t) = \sin^2\theta(t)\dot{\varphi}(t)/2$ and $\Omega(t)$ $=$ ${\rm i} e^{{\rm i}\varphi(t)} [\dot{\theta}(t)$ $+{\rm i}\sin\theta(t)\cos\theta(t)\dot{\varphi}(t)]/2$.
In our scheme, we choose a time-dependent detuning $\Delta(t)$ to more precisely construct on-demand trajectories and optimize the performance of geometric quantum gates, which is within current state-of-the-art technologies  \cite{Leung2018,Landsman2019,Roth2017,PHY6}.
Because $\langle\psi_{k}(t)|H(t)|\psi_{k}(t)\rangle=0$, the evolution states $|\psi_{k}(t)\rangle$ will only accumulate geometric phases, i.e., $\gamma_{k}(t)={\rm i}\int_{0}^{t}\langle\mu_{k}(t')|\dot{\mu}_{k}(t')\rangle dt'$.
When the auxiliary vectors meet the cyclic evolution for a duration of $\tau$, i.e., $|\mu_k(\tau)\rangle=|\mu_k(0)\rangle=|\psi_{k}(0)\rangle$, the corresponding evolution operator will be
\begin{eqnarray}
\label{operator}
U(\tau) = \!\!\!\sum^2_{k=1}|\psi_{k}(\tau)\rangle\langle \psi_{k}(0)| 
= \!\!\!\sum^2_{k=1}e^{{\rm i}\gamma_{k}(\tau)}|\mu_k(0)\rangle\langle\mu_{k}(0)|.
\end{eqnarray}
%where $\gamma_1(\tau)=-\gamma_2(\tau)= -\frac{1}{2}\int_0^{\tau}[1-\cos\theta(t)]\dot{\varphi}(t)dt$.
By setting $\theta_0$ = $\theta(0)$ and $\varphi_0$ = $\varphi(0)$, the evolution operator, in the computational space spanned by $\{|0\rangle,|1\rangle\}$, will reduce to
\begin{eqnarray} \label{gate}
U(\tau)=\exp\left(-{\rm i} {\gamma \over 2} \mathbf{n} \cdot \mathbf{\sigma}\right),
\end{eqnarray}
where $\mathbf{n}=(\sin\theta_0\cos\varphi_0, \sin\theta_0\sin\varphi_0, \cos\theta_0)$ is a unit directional  vector and $\mathbf{\sigma}=(\sigma_x, \sigma_y, \sigma_z)$ is a vector of Pauli operators. Clearly, the  operator $U(\tau)$ represents an arbitrary rotation gate around the  axis $\mathbf{n}$ by an  angle $\gamma$, with the geometric phase being
\begin{eqnarray}
\label{phase}
\gamma=  \int_0^{\tau}[1-\cos\theta(t)]\dot{\varphi}(t)dt.
\end{eqnarray}
Since $(\theta(t),\varphi(t))$ defines a point on the Bloch sphere, this trajectory  forms a closed path $C$ on the Bloch sphere during the construction of a geometric quantum gate. In particular, for a given $\gamma$, different choices of $\theta(t)$ and $\varphi(t)$ will determine different trajectories of the geometric evolution. Thus, generally,  there are infinite options for implementing a certain geometric gate. In addition, to be experimentally friendly, i.e., the time dependence of the amplitude of a pulse is set to change from zero to zero during the gate’s implementation, generally,  $\nu(t) \in \{\theta(t), \varphi(t)\}$ can be set as
\begin{eqnarray}
\label{fuliye}
\nu(t)= \mathcal{D}^\nu + \sum_{n=1}^N a_n^\nu \sin\left(\frac{b_n^\nu \pi t}{\tau}\right)^{c_n^\nu},
%&&\theta(t)=\mathcal{A}+a_1\sin(\frac{b_1\pi t}{\tau})^{c_1}+...+a_n\sin(\frac{b_n\pi t}{\tau})^{c_n}+..., \notag \\
%&&\varphi(t)=\mathcal{D}+d_1\sin(\frac{e_1\pi t}{\tau})^{f_1}+...+d_n\sin(\frac{e_n\pi t}{\tau})^{f_n}+....\notag \\
\end{eqnarray}
where $a_n^\nu, b_n^\nu$, and $c_n^\nu$ are the free parameters, and $\mathcal{D}^\nu$ is a free function that determines the initial values and shapes of $\theta(t)$ and $\varphi(t)$. As shown in Eq (4), the type of geometric quantum gate is determined by the initial values of $\theta(t)$ and $\varphi(t)$. Since different quantum gates require different initial values for $\theta(t)$ and $\varphi(t)$, the corresponding free parameter $\mathcal{D}$ also varies. With  $\theta(t)$ and $\varphi(t)$ being set, we can calculate the obtained geometric phase.  In our construction, typical quantum gates can be obtained  with smooth pulses instead of being divided into segments as in conventional single-loop schemes \cite{loop,loop4,loop6}. 
Furthermore, since only the area of the trajectory influences specific geometric quantum gates, we have the flexibility to alter the geometric shape of the trajectory in different scenarios. This eliminates the need to divide the evolution process into orange-slice paths, as illustrated in Figs. \ref{path}(a) and \ref{path}(b), in the conventional single-loop NGQG scheme (see the Appendix for the implementation details).

\begin{table*}[tbp]
\centering
\begin{tabular}{ccccccccccccccc} %
\hline
\hline
Cases\qquad\qquad&       Gate \qquad\qquad&  $a_1^{\theta}$\quad \quad& $a_2^{\theta}$\quad \quad& $a_3^{\theta}$\quad \quad & $a_4^{\theta}$\quad \quad& $a_1^{\varphi}$\quad \quad & $a_2^{\varphi}$\quad \quad& $a_3^{\varphi}$\quad \quad & $a_4^{\varphi}$\quad \quad&  \quad \quad  $S/\pi$\\
\hline
\multirow{2}*{1}\qquad\qquad&      $T$ \qquad\qquad&   -1.03\quad \quad& -0.11\quad \quad& 0.02\quad \quad& 0.01\quad \quad& 2.16\quad \quad& 0.59\quad \quad& 0.61\quad \quad& -0.69\quad \quad& \quad \quad0.404\\

                               &      $H$\qquad\qquad&  0.44\quad \quad& -0.064\quad \quad& -0.01\quad \quad& 0.01\quad \quad& 0\quad \quad& 0\quad \quad& 0\quad \quad&0\quad \quad& \quad \quad0.441\\
 \\
\multirow{2}*{2}\qquad\qquad&      $T$ \qquad\qquad&  0.91\quad \quad& 0.1\quad \quad& 0.03\quad \quad& 0.016\quad \quad& -2.76\quad \quad& 4.83\quad \quad& 10\quad \quad& -9.65\quad \quad& \quad \quad0.408\\

                               &      $H$\qquad\qquad&  0.41\quad \quad& 0.01\quad \quad& -0.03\quad \quad& -0.01\quad \quad& 0\quad \quad& 0\quad \quad& 0\quad \quad& 0\quad \quad& \quad \quad0.444 \\
\\
\multirow{2}*{3}\qquad\qquad&   $T$ \qquad\qquad&   -1.87\quad \quad& 0.57\quad \quad& 0.1\quad \quad& -0.12\quad \quad& 0.24\quad \quad& -0.15\quad \quad& 0.33\quad \quad& 1.1\quad \quad&\quad \quad0.569\\

                                            &   $H$\qquad\qquad& 1.164\quad \quad& -0.46\quad \quad& -0.524\quad \quad& -0.17\quad \quad& 0\quad \quad& 0\quad \quad& 0\quad \quad& 0\quad \quad&\quad \quad0.754 \\
 \\
\multirow{2}*{4}\qquad\qquad&   $T$ \qquad\qquad&   -4.61\quad \quad& 0\quad \quad& -0.23\quad \quad& 0\quad \quad& -2.14\quad \quad& 2.345\quad \quad& 5.84\quad \quad& -5.58\quad \quad& \quad \quad1.541 \\

                                            & $H$\qquad\qquad&  0.413\quad \quad&	0.02\quad \quad& -0.035\quad \quad& -0.017\quad \quad& 0\quad \quad& 0\quad \quad& 0\quad \quad& 0\quad \quad& \quad \quad 0.445\\
\\
\multirow{3}*{5}\qquad\qquad&   $T$ \qquad\qquad&   -4.62\quad \quad& -0.02\quad \quad& -0.25\quad \quad&-0.01\quad \quad& 6.305\quad \quad& -8.85\quad \quad&-2.23\quad \quad&5.57\quad \quad& \quad \quad1.551  \\

                                          &   $H$\qquad\qquad&  0.38\quad \quad& 0.02\quad \quad& -0.012\quad \quad& 0\quad \quad& 0\quad \quad& 0\quad \quad& 0\quad \quad& 0\quad \quad& \quad \quad0.445  \\
 &   $CP$\qquad\qquad&  -1.265\quad \quad& -0.178\quad \quad& -0.09\quad \quad& -0.027\quad \quad& 4.388\quad \quad& 2.738\quad \quad& 0.154\quad \quad& -1.079\quad \quad& \quad \quad-  \\
\hline
\hline
\end{tabular}
\caption{Optimized  parameters that determine  pulse shapes for ONGQG scheme under different scenarios.}
\label{Table1}
\end{table*}

In the following, taking the $T$ and Hadamard ($H$) gates as examples, we optimize geometric quantum gates under various scenarios. The initial parameters $(\theta_0, \varphi_0, \gamma)$ for the $T$ and  $H$ gates correspond to $(0, 0, \pi/8)$ and $(\pi/4, 0, \pi/2)$, respectively. To expedite the numerical search process for the parameters in Eq. (\ref{fuliye}), we set the parameters $ \mathcal{D}^\nu$, $b_n^\nu$, and $c_n^\nu$ in a fixed configuration. Specifically, when constructing the $T$ gate, these parameters are set to $ \mathcal{D}^\nu=0$, $b_n^\theta=n$, $b_n^\varphi=1/2$, $c_n^\theta=2$, and $c_n^\varphi=n+1$. For the construction of the $H$ gate, we configure them as $ \mathcal{D}^\theta=\pi/4$, $ \mathcal{D}^\varphi=2\pi\sin^2[\pi t/(2\tau)]$, $b_n^\nu=n$, $c_n^\nu=2$. Consequently, our primary focus is on optimizing the parameters $a_n^\nu$, and the corresponding optimized parameters are detailed in Table \ref{Table1}. In addition, we truncate at $n=4$ here; this is arbitrary, but this setting is accurate enough for our purpose.

\section{Gate optimization}
In this section, we illustrate the flexibility of our scheme by optimizing it under different practical scenarios.

\subsection{Fidelity optimization}

The main drawback of the conventional single-loop NGQG schemes is that the pulse areas required for constructing different geometric quantum gates are the same, and this area is usually at least twice that of the corresponding dynamical quantum gates. Thus, geometric quantum gates are usually more time and energy consuming, which will lead to lower gate fidelity due to the decoherence effect, despite their enhancement in terms of gate robustness. Therefore, our first example, which we term as Case 1, is to optimize the pulse area for geometric quantum gates by trajectory design, i.e., optimizing the parameters in $\nu(t)$. Here, we define the pulse area as
\begin{eqnarray}
\label{area}
S&=& \int_0^{\tau}\Omega'(t)dt \notag \\
&=&\int_0^{\tau} {1\over 2}\sqrt{\left[\dot{\varphi}(t)\sin\theta(t)\cos\theta(t)\right]^{2}+\dot{\theta}^{2}(t)} dt, %\notag
\end{eqnarray}
where $\Omega'(t)= \sqrt{\left[\dot{\varphi}(t)\sin\theta(t)\cos\theta(t)\right]^{2}+\dot{\theta}^{2}(t)}/2$ represents the amplitude part of the Rabi frequency $\Omega(t)$ defined in Eq. (2). Our target is to minimize  $S$ for a certain geometric gate.

\begin{figure}[tbp]
  \centering
\includegraphics[width=1 \linewidth]{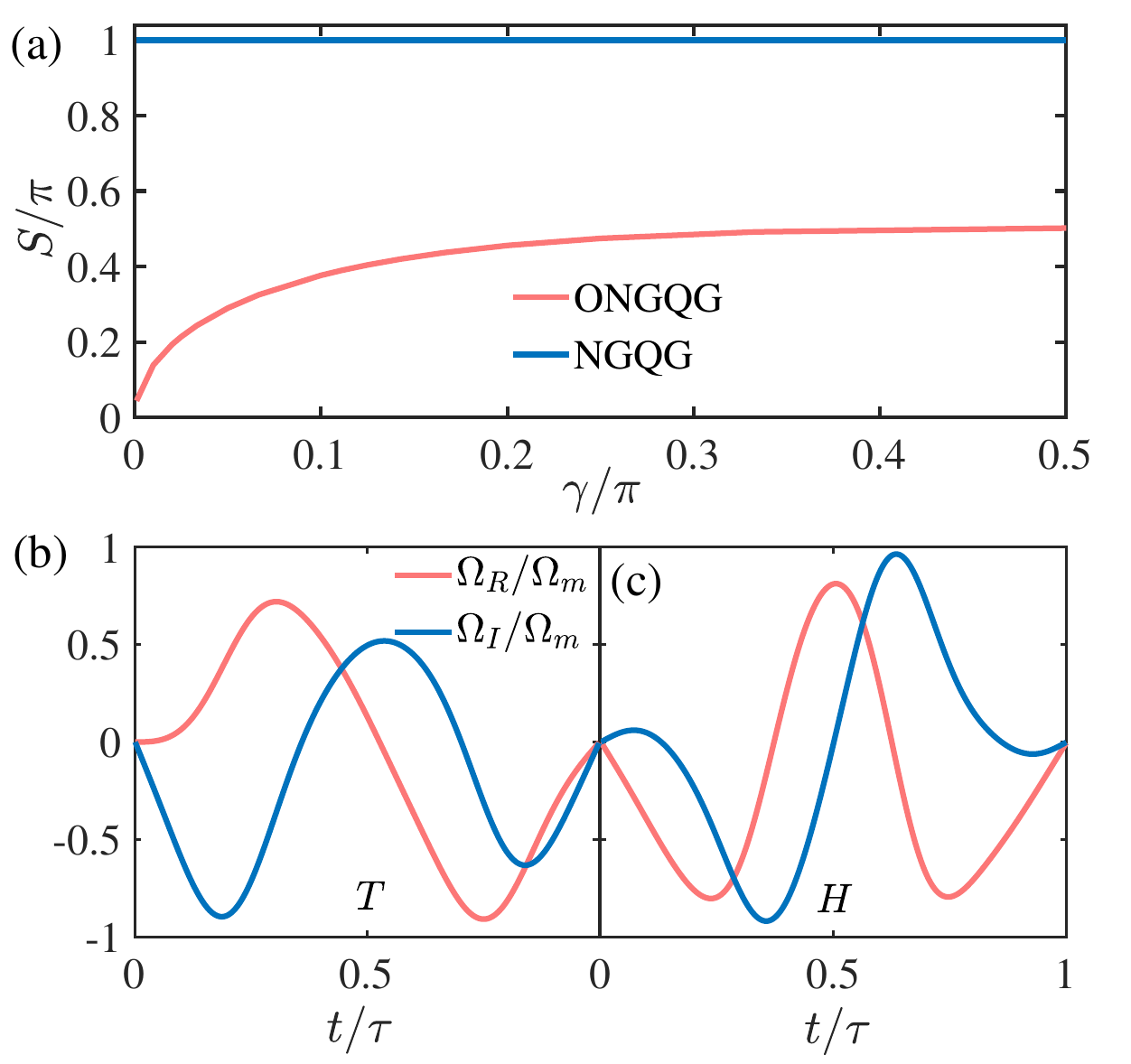}
\caption{Results for the pulse-area optimization of our scheme.
(a) Pulse area of $Z$ rotation with different angles $\gamma$ of our scheme
and the single-loop NGQG scheme. (b), (c) Pulse shapes with
minimal pulse area for the $T$ and $H$ gates, respectively}
 \label{figpulse}
 \end{figure}

\begin{figure}[t]
  \centering
  \includegraphics[width=1\linewidth]{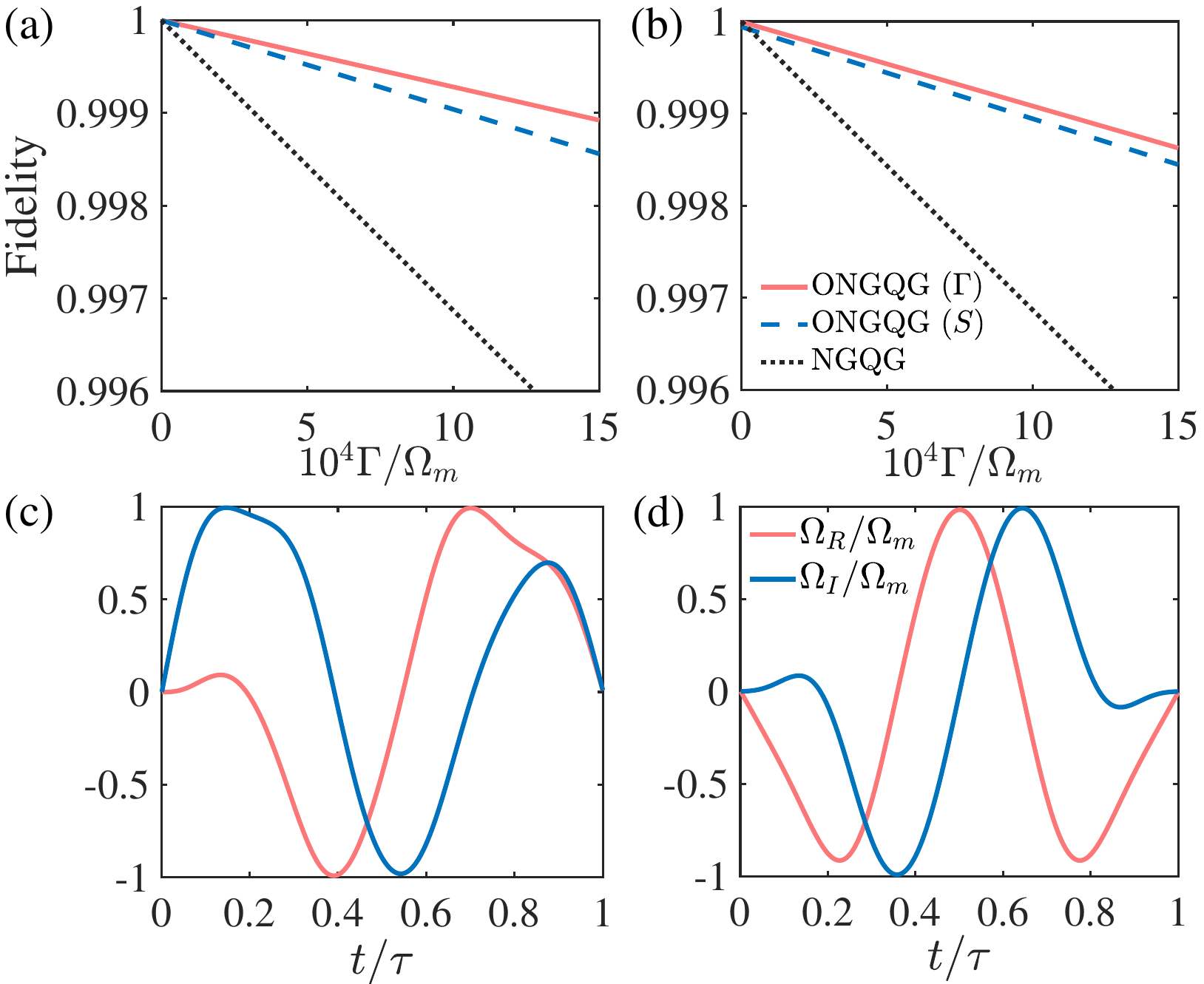}
  \caption{ Results of the fidelity optimization. Comparison of
performance under decoherence in different schemes for (a) $T$
gate and (b) $H$ gate, in which the red solid, blue dashed, and
black dotted lines represent the fidelity-optimized, pulse-area
optimized, and single-loop NGQG schemes, respectively. (c),(d)
Real and imaginary parts of the pulses for the $T$ and $H$ gates at
the highest fidelity. }
  \label{decay}
\end{figure}

In Fig. \ref{figpulse}(a), we plot the minimized pulse area $S$ for a $Z$-rotation gate with geometric phase of $\gamma$; for a conventional single-loop NGQG, $S$ is a fix value of $\pi$, no matter how small  $\gamma$ is. In contrast, in our optimized NGQG (ONGQG) scheme, the pulse areas for the $Z$-axis rotation gates  $S_z \leq  \pi/2$, and $S_z$ decreases as $\gamma$ decreases. As an example, for the $T$ gate, its pulse area $S_T\approx 0.404\pi$; 
 the corresponding trajectory is plotted in Fig. \ref{path}(c).  This shows that our construction can greatly minimize time and energy consumption, leading to higher-quality geometric gates.
We find that for an $H$ gate, which has the maximum rotation angle, $S_H$ is still less than $\pi/2$, as listed in table \ref{Table1}. The corresponding trajectory is plotted in Fig. \ref{path}(d), this is approximately a circle. Under the optimized pulse area, the smooth pulse shapes for the $T$  and $H$ gates are plotted in Figs. \ref{figpulse}(b) and \ref{figpulse}(c), respectively. Hhere 
$\Omega_R=[-\dot{\varphi}(t)\sin\theta(t)\cos\theta(t)\cos\varphi(t)-\dot{\theta}(t)\sin\varphi(t)]/2$ and $\Omega_I=[-\dot{\varphi}(t)\sin\theta(t)\cos\theta(t)\sin\varphi(t)+\dot{\theta}(t)\cos\varphi(t)]/2$ are the respective real and imaginary parts of $\Omega(t)$ in Eq. (\ref{hamiltonian}), and $\Omega_m$ is the maximum value of $\Omega(t)$. Due to the decrease of the pulse area, the infidelity of the two gates will also be decreased compared with that of the single-loop NGQG scheme, as shown in Figs. \ref{decay}(a) and \ref{decay}(b).

%\subsection{Fidelity optimization}

For Case 2, to fully evaluate the detrimental effect of the decoherence in constructing an NGQG, we use Lindblad master equation of Ref. \cite{master}
\begin{eqnarray}
\label{EqMaster}
\dot\rho(t)&=&-{\rm i}[H(t), \rho(t)]+\frac {1} {2}\sum_{j=-,z}\Gamma_{j}L(\sigma_{j}),
\end{eqnarray}
where $\rho(t)$ is the density matrix of the quantum system, and $L(A)=2A\rho A-A^{\dag}A\rho-\rho A^{\dag}A $ is the Lindbladian operator with $\sigma_-=|0\rangle\langle 1| $ and $\sigma_z=(|1\rangle\langle 1| -|0\rangle\langle0|)/2$, $\Gamma_-$ and $\Gamma_z$ represent the decay and dephasing rates, respectively, and we set  $\Gamma_-=\Gamma_z=\Gamma$ for demonstration purpose. The gate fidelity is defined as
\begin{eqnarray}
\label{Fidelity}
F=  {1 \over 6} \sum_{l=1}^6\langle\Psi_l(0)|U(\tau)^{\dag}\rho U(\tau)|\Psi_l(0)\rangle,
\end{eqnarray}
where the six initial states $|\Psi_l(0)\rangle$ are $|0\rangle$, $|1\rangle$, $(|0\rangle+|1\rangle)/\sqrt{2}$, $(|0\rangle-|1\rangle)/\sqrt{2}$, $(|0\rangle+{\rm i}|1\rangle)/\sqrt{2}$, and $(|0\rangle-{\rm i}|1\rangle)/\sqrt{2}$; these have been used to  evaluate the performance of quantum gate \cite{Schwinger1960,IvanoviC1981,Klappenecker2005}. Under the optimized parameters, as listed in the Table \ref{Table1}, we find that the resistance of $T$ and $H$ gates to decoherence in our scheme is considerately improved in comparison with the single-loop NGQG scheme; this is shown in Figs. \ref{decay}(a) and \ref{decay}(b), and the corresponding pulse shapes are presented in Figs. \ref{decay}(c) and \ref{decay}(d), respectively. The corresponding optimized trajectories are shown in Figs. \ref{path}(e) and \ref{path}(f), respectively.
The gate fidelity is over 99.99\% for decoherence rates lower than $\Omega_m/10000$, which is now achievable in various quantum systems \cite{Muhonen2014,Ballance2016,Somoroff2023}. In addition, this also indicates that minimizing the pulse area does not correspond to achieving the maximum fidelity.

 \begin{figure}[t]
  \centering
  \includegraphics[width=1\linewidth]{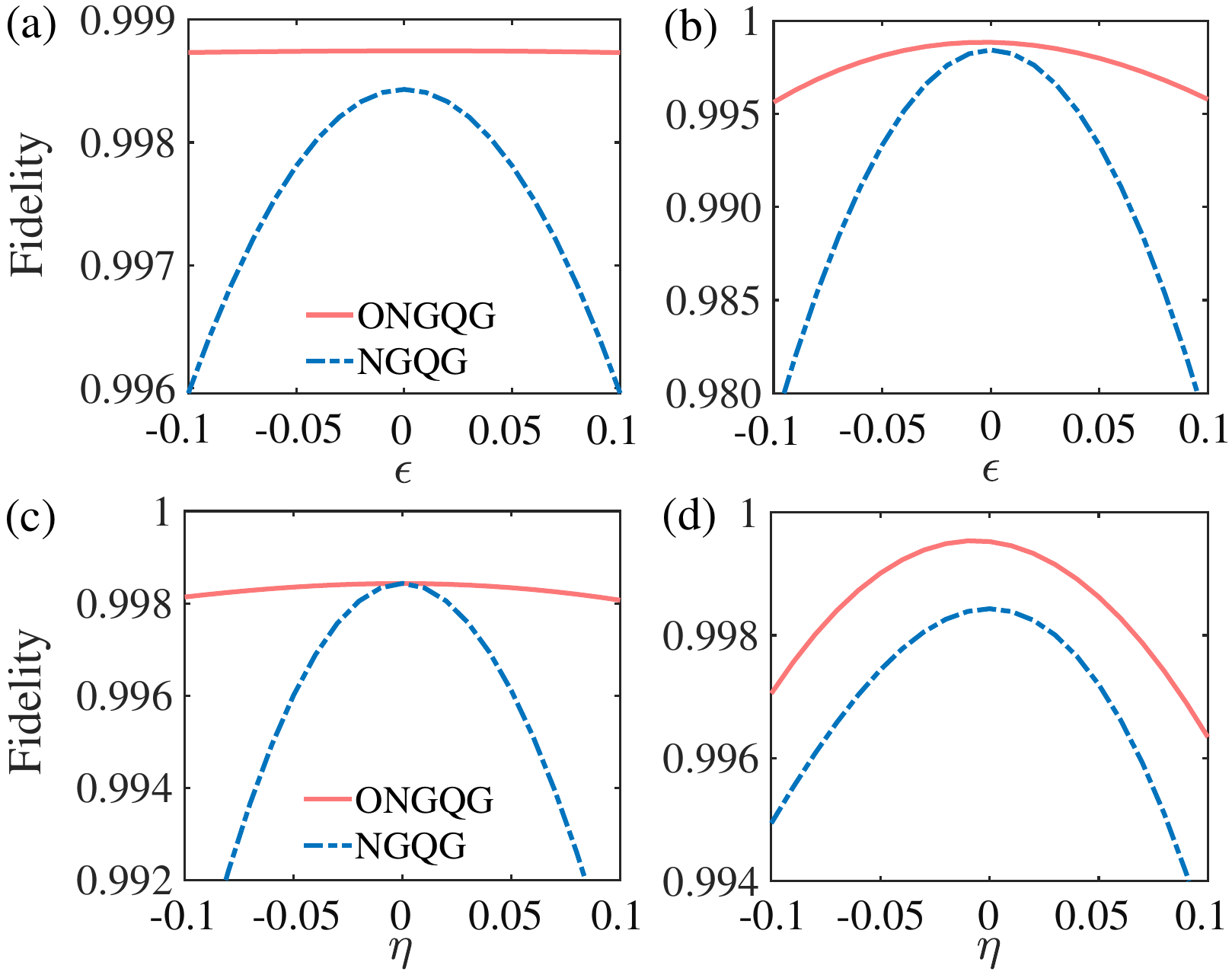}
\caption{Gate performance with errors and under decoherence rate $\Gamma=\Omega_m/2000$. (a)   $T$  and (b) $H$ gates  with  $\epsilon$ error; (c) $T$  and  (d) $H$ gates with   $\eta$   error. }
  \label{error}
\end{figure}

%\newpage
\subsection{Robustness optimization}

The other key indicator for the quality of a quantum operation is its robustness. For a quantum system described by the Hamiltonian $H(t)$ in Eq. (\ref{hamiltonian}), the gate robustness can be evaluated by the gate fidelity under operational errors $V(t)=\epsilon \Omega(t) (|0\rangle\langle 1|+\rm{H.c.})+ \eta\Omega_m(|0\rangle\langle 0|-|1\rangle\langle 1|)/2$, where $\epsilon$  and $\eta$ represent the Rabi frequency error and the detuning error, which are induced by various errors in the $\sigma_x$ and $\sigma_z$ directions. The gate fidelity is defined as \cite{XGWang2009}
\begin{eqnarray}
\label{gfidelity}
F&=&\frac{|{\rm Tr} [U_{\epsilon}^{\dag}(\tau)U(\tau)]|}{|{\rm Tr} [U^{\dag}(\tau)U(\tau)]|},\notag \\
&\approx& 1-\frac{1}{2}\left|\int_0^{\tau}e(t)dt\right|^2-\frac{1}{2}\left|\int_0^{\tau}g(t)dt\right|^2,
\end{eqnarray}
where $U(\tau)$ (as shown in Eq. (\ref{operator})) and $U_{\epsilon}(\tau)=\sum_{i=1}^2|\psi^{\epsilon}_i(\tau)\rangle\langle\psi_i(0)|$ are the corresponding evolution operator in the absence and presence of the quantum error $V(t)$, respectively, with $|\psi^{\epsilon}_i (t)$ being the evolution state under the Hamiltonian  $H_{\epsilon}(t)= H(t)+V(t)$.  In addition, \cite{shot2013}
\begin{subequations}
\begin{align}
&g(t)=\langle \psi_{1}(t)|V(t)|\psi_{2}(t)\rangle=g^{\epsilon}(t)+g^{\eta}(t),  \\
&e(t)=\langle \psi_{1}(t)|V(t)|\psi_{1}(t)\rangle=e^{\epsilon}(t)+e^{\eta}(t),
\end{align}
\end{subequations}
with $g^{\epsilon}(t)= \epsilon  [\dot{\varphi}(t)\sin\theta(t)\cos^2\theta(t)+{\rm i}\dot{\theta}(t)]\exp\{-{\rm i}[\varphi(t)+2\gamma(t)]\}/2$,
$g^{\eta}(t)=\eta\Omega_m \sin\theta(t) \exp\{-{\rm i}[\varphi(t)+2\gamma(t)]\}/2, 
e^{\epsilon}(t)=- \epsilon\dot{\varphi}(t)\sin^2\theta(t)\cos\theta(t)/2,$
and $e^{\eta}(t)=\eta\Omega_m \cos\theta(t)/2$.
For a certain gate with geometric phase $\gamma$, we define the cost function, i.e., the gate infidelity, as
\begin{eqnarray}
\label{costfunrabi}
\mathcal{F}^{\xi}(a_n^{\nu})&=& \left |\int_0^{\tau}e^{\xi}(t)dt \right| + \left |\int_0^{\tau}{\rm Re}(g^{\xi}(t))dt\right|\notag \\
&&+ \left |\int_0^{\tau}{\rm Im}(g^{\xi}(t))dt\right|,
\end{eqnarray}
for $\xi$ ($\xi=\epsilon,\ \eta$) error, where ${\rm Re}(g^{\xi}(t))$ and ${\rm Im}(g^{\xi}(t))$ are the real and imaginary parts of $g^{\xi}(t)$, respectively.

 \begin{figure}[tbp]
\centering
\includegraphics[width=0.9\linewidth]{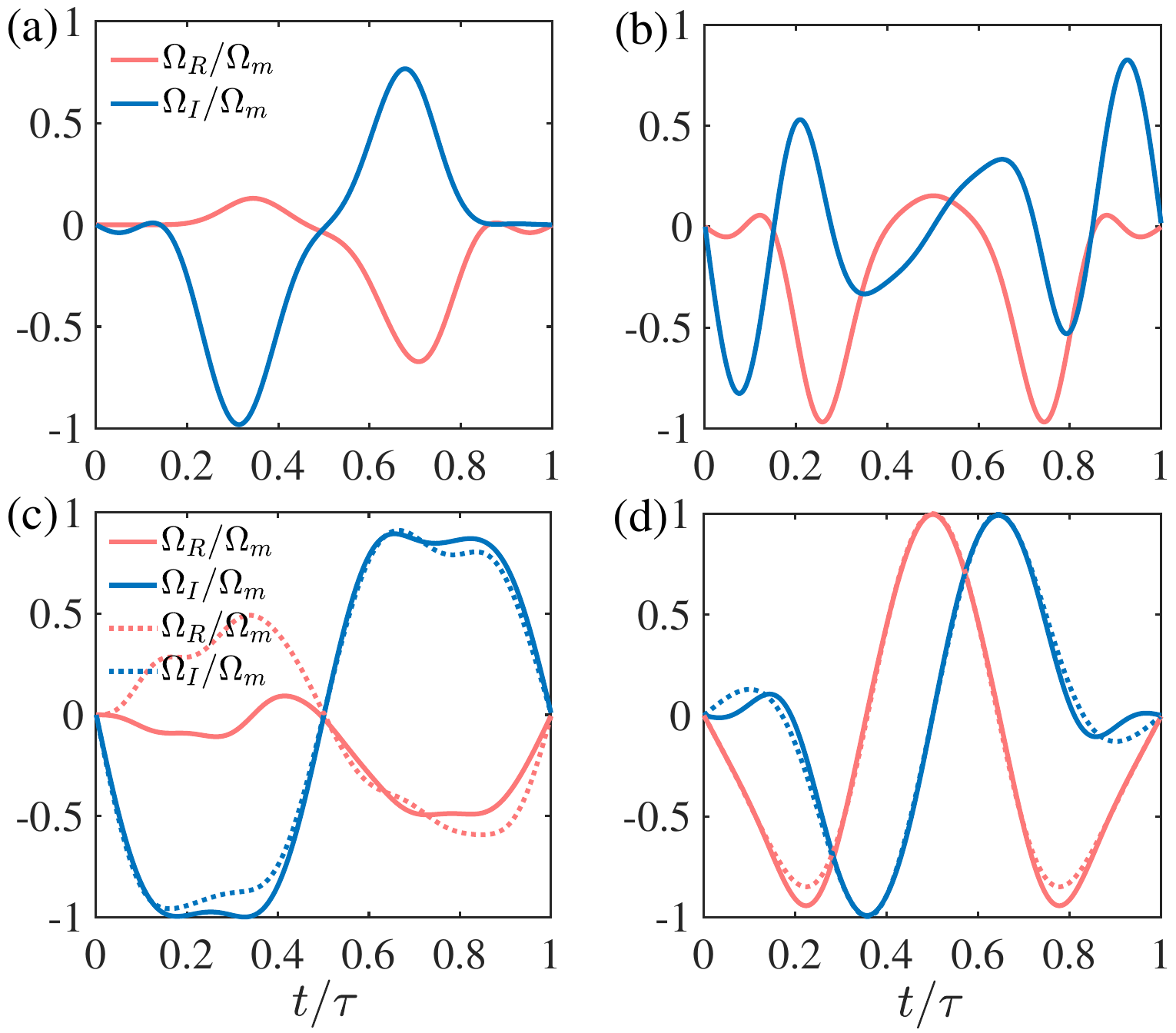}
\caption{Optimized pulse shapes. (a), (b) Real and imaginary
parts of the pulses for the $T$ and $H$ gates in Case 3, respectively.
(c), (d) Real and imaginary parts of the pulse shapes for the $T$ and
$H$ gates, with solid lines representing Case 4 and dashed lines
representing Case 5.}
  \label{pulse}
\end{figure}

Next, we optimize the gate robustness in the presence of different quantum errors, i.e., the $\epsilon$ or/and $\eta$ errors, under the decoherence effect, with the unified decoherence rate being $\Gamma=\Omega_m/2000$, for demonstration purpose. We numerically find the parameters $a_n^{\nu}$ to minimize the cost function, as listed in table \ref{Table1}. For Case 3, i.e., the  $\epsilon$  error case,  the results are shown in FigS. \ref{error}(a) and  \ref{error}(b) for the $T$ and $H$ gates, respectively. For Case 4, i.e., the $\eta$ error cases,  the results are shown in Figs. \ref{error}(c) and \ref{error}(d)  for the $T$ and $H$ gate, respectively. The corresponding
optimized trajectories are shown in Figs. \ref{path}(g)-(j), respectively, and the corresponding pulse shapes are plotted in Fig. \ref{pulse}. 
In both cases, our ONGQG scheme demonstrates considerate enhancement in terms of gate robustness. In Fig. \ref{error}, we also compare the results  of our scheme with the conventional single-loop NGQG scheme; this comparison indicates that the  gate robustness in our ONGQG scheme is superior to that in the single-loop scheme. 

\begin{figure}[tbp]
\centering
\includegraphics[width=1\linewidth]{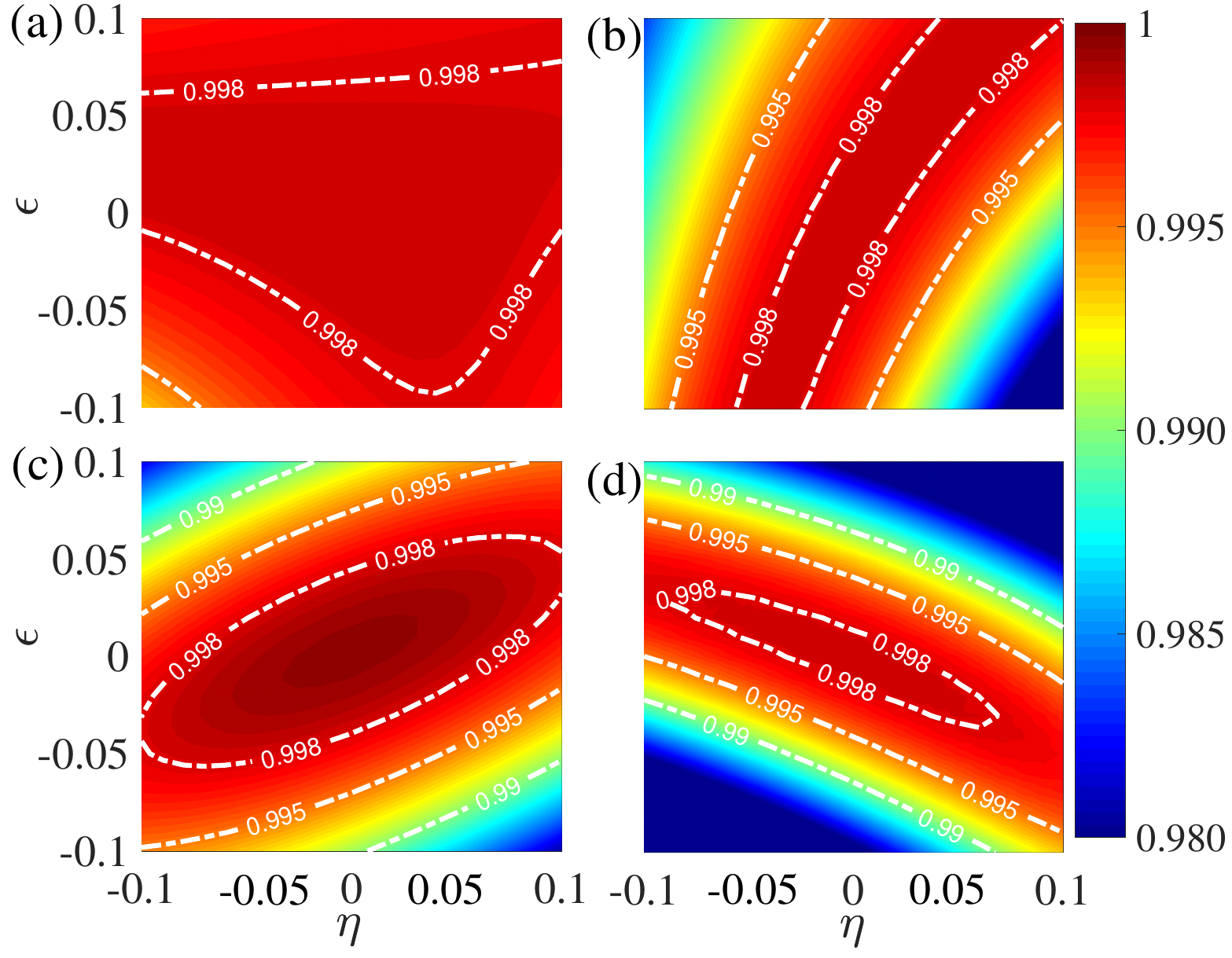}
\caption{Gate performance with both  $\epsilon$  and  $\eta$  errors and under decoherence rate $\Gamma=\Omega_m/2000$. (a), (b) Results for $T$ gates with
our scheme and the single-loop NGQG scheme, respectively.
(c), (d) Results for $H$ gates with our scheme and the single-loop
NGQG scheme, respectively. }
  \label{figtwoerror}
\end{figure}

Furthermore, for Case 5, we consider both type of errors under the decoherence effect by using the cost function $$\mathcal{F}^{\epsilon,\eta}(a_n^{\nu})=\mathcal{F}^{\epsilon} (a_n^{\nu})+\mathcal{F}^{\eta}(a_n^{\nu}).$$ In this case, the optimized trajectories are shown in Figs. \ref{path}(i) and (j), and the corresponding pulse shapes are plotted in Figs. \ref{pulse}(c) and (d). As shown in Fig. \ref{figtwoerror}, we can still find optimized parameters that can  strengthen the gate robustness against both errors, indicating that it is a  promising protocol  for practical quantum computation.

\section{ Physical implementation}\label{99}

In this section, we describe the implementation of our ONGQG scheme in a superconducting quantum circuit consisting of capacitively coupled transmon qubits, and we evaluate the performance of geometric single-qubit and nontrivial two-qubit gates under current state-of-the-art experimental parameters.

\subsection{Universal single-qubit gate}

\begin{figure}[t]  
  \centering
  \includegraphics[width=1\linewidth]{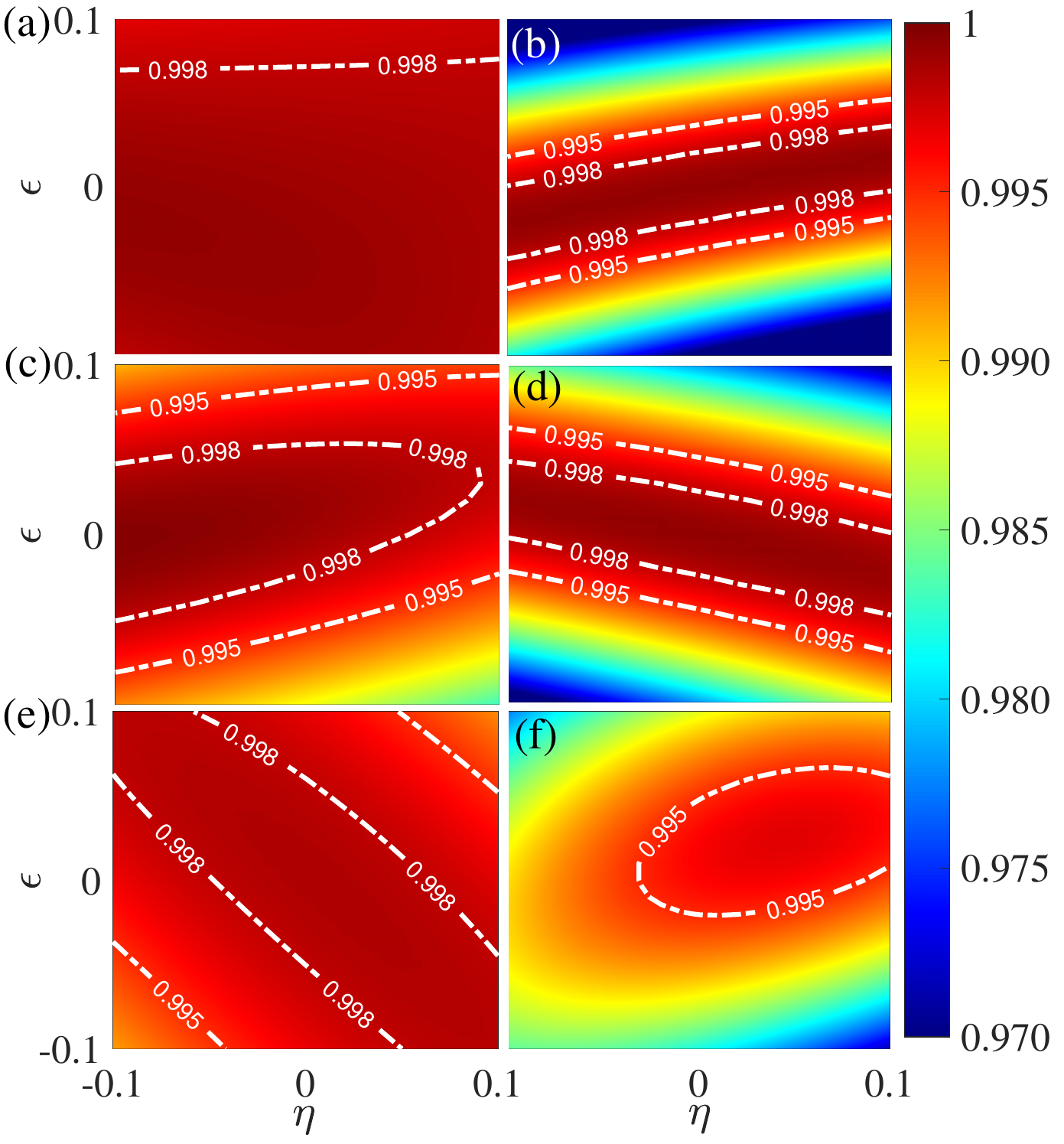}
  \caption{The performance of quantum gates implemented in superconducting quantum systems  with both Rabi and detuning errors and  under decoherence rates $\Gamma_-=\Gamma_z=2\pi\times2$ kHz, showing:
the $T$ gate in the (a) ONGQG and (b) NGQG schemes; the $H$ gate
in the (c) ONGQG and (d) NGQG schemes; and the $CP$ gate in
the (e) ONGQG and (f) NGQG schemes. } 
  \label{fig12}
\end{figure}

We first implement the single-qubit gates. The two lowest energy levels of a transmon qubit, denoted as $\{|0\rangle, |1\rangle\}$, are employed as our computational subspace; however, due to the weak anharmonicity of the transmon qubit, driving the computational subspace inevitably induces simultaneous coupling to higher energy states, leading to information leakage into $|2\rangle$ or higher levels. Here, we only take into account the qubit-state leakage to the state $|2\rangle$, which represents the primary leakage error for transmon qubits. To suppress this leakage, we implement the “derivative removal by adiabatic gate" (DRAG) technique \cite{PHY3,PHY4}. The Hamiltonian of the quantum system, modified using the DRAG technique, can be expressed as 
\begin{eqnarray}
\label{67}
\mathcal{H}_{D}(t)&=&\frac{1}{2}[\mathbf{B}_{0}(t)+\mathbf{B}_{d}(t)]\cdot \mathbf{S}-\alpha|2\rangle\langle2|,
\end{eqnarray}
where $\alpha$ is the anharmonicity of the transmon qubit, and $\mathbf{B}_0(t)$ and $\mathbf{B}_d(t)$ represent the original microwave drive and the DRAG-corrected microwave drive, respectively. The specific expressions are given by 
\begin{eqnarray}
\label{52}
\mathbf{B}_{0}(t)&=&
\begin{cases}
B_{x}(t) =\Omega_R(t), \\
    
    B_y(t) =\Omega_I(t),  \\     
    B_z(t) =-\Delta(t),\\
\end{cases} \\
\mathbf{B}_{d}(t)&=&
\begin{cases}
B_{dx}(t) =\frac{1}{2\alpha}\left(\dot{B}_y(t)-B_z(t)B_x(t)\right), \\
    
    B_{dy}(t) =-\frac{1}{2\alpha}\left(\dot{B}_x(t)+B_z(t)B_y(t)\right),  \\     
    B_{dz}(t) =0,\\
\end{cases}
\end{eqnarray}
where $\Omega_R(t)$ and $\Omega_I(t)$ correspond to the real and imaginary parts of the driving field $\Omega(t)$, respectively. In addition, the components of the vector $\mathbf{S}$ in the $x$, $y$, and $z$ directions are 
\begin{eqnarray}
\label{53}
\mathbf{S}&=&
\begin{cases}
S_{x} =\sum\limits_{b=1,2} \sqrt{b}\left(|b\rangle\langle b-1|+|b-1\rangle\langle b|\right), \\
    
    S_y =\sum\limits_{b=1,2}\sqrt{b}\left(\rm{i}|b\rangle\langle b-1|-\rm{i}|b-1\rangle\langle b|\right),  \\     
    S_z =\sum\limits_{b=1,2,3}(3-2b)|b-1\rangle\langle b-1|.
\end{cases}
\end{eqnarray}
Next, we continue to assess the performance of quantum gates in a superconducting transmon qubit system, using the $H$ and $T$ gates as illustrative examples. In this context, the decoherence operators in the master equation take the form $\sigma_{-}=|0\rangle\langle1|+\sqrt{2}|1\rangle\langle2|$ and $\sigma_{z}=|1\rangle\langle1|+2|2\rangle\langle2|$, with the decoherence rates being $\Gamma_-=\Gamma_z=2\pi\times2$ kHz. In our simulations, we set the maximum amplitude to $\Omega_m=2\pi\times 10$ MHz and choose an anharmonicity of $\alpha=2\pi\times 300$ MHz, which is within experimentally achievable limits. The Rabi error is modeled as $\Omega(t)\rightarrow(1+\epsilon)\Omega(t)$, and the detuning error is $\eta\Omega_m(|1\rangle\langle1|+2|2\rangle\langle2|)/2$. Using the optimized parameters from Case 5 in Table \ref{Table1}, we depict the comprehensive performance of the $H$ and $T$ gates in Figs. \ref{fig12}(a) and \ref{fig12}(c), respectively. When compared with the conventional NGQG approach, depicted in Figs. \ref{fig12}(b) and \ref{fig12}(d), our method demonstrates significant improvement.

\subsection{ Nontrivial two-qubit geometric gate}
We next proceed to implement nontrivial two-qubit geometric gates with two capacitively coupled transmon qubits. The Hamiltonian is  
\begin{equation}
\label{54}
\begin{split}
\mathcal{H}_{12}=&\sum_{i=1}^{2}\sum_{j=1}^{\infty}[i\omega_i-\frac{j(j-1)}{2}\alpha_i]|j\rangle_i\langle j| \\
+&g_{12}[(\sum_{p=0}^{\infty}\sqrt{p+1}|p\rangle_1\langle p+1|\\
\otimes&\sum_{q=0}^{\infty}\sqrt{q+1}|q\rangle_2\langle q+1|)+\rm{H.c.}],
\end{split}
\end{equation}
where $g_{12}$ represents the fixed coupling strength between the two transmon qubits, and $\omega_i$ and $\alpha_i$ are the inherent frequency and anharmonicity of qubit $i$. To achieve tunable coupling \cite{PHY6,PHY7,Roth2017}, an ac frequency drive is applied to transmon qubit 2 to modulate its qubit frequency, i.e., $\omega_2(t)=\omega_2+\dot{f}(t)$, where $f(t)=\beta\sin\left[\int_0^t\nu(t')dt'+\chi(t)\right]$. 
In the interaction picture, truncating the Hamiltonian to the single- and double-excitation subspaces, Eq. (\ref{54}) reduces to  
\begin{eqnarray}
\label{55}
\mathcal{H}'_{12}(t)& =&g_{12}\{[|10\rangle\langle01|e^{{{\rm i}}\Delta_{12}t}+\sqrt{2}|11\rangle\langle02|e^{{\rm i}(\Delta_{12}+\alpha_2)t} \notag\\
&& + \sqrt{2}|20\rangle\langle11|e^{{\rm i}(\Delta_{12}-\alpha_1)t}]e^{-{\rm i}\beta\sin[\int_0^t\nu(t')dt'+\chi(t)]}\}\notag\\
 &&+ {\rm{H.c.}},
\end{eqnarray}
where $\Delta_{12}=\omega_1-\omega_2$. This  truncation is ensured by the fact that $|2\rangle$ is the main leakage source for the qubit states. Utilizing the Jacobi–Anger identity ${\rm exp}({\rm i}\beta\cos\theta)=\sum_{n=-\infty}^{+\infty}{\rm i}^{n}J_n(\beta){\rm exp}({\rm i}n\theta)$ and making a representation transformation with unitary operator $U_{\Delta'}={\rm exp}\left[-{\rm i}\int_0^tdt'\Delta'(t')(|02\rangle\langle02|-|11\rangle\langle11|)/2\right]$, we obtain, after setting $\nu(t)=\Delta_{12}+\Delta'(t)+\alpha_2$ and neglecting high-frequency oscillatory terms, the Hamiltonian in the two-qubit subspace $\{|02\rangle,\ |11\rangle\}$ as
\begin{equation}
\label{56}
\begin{split}
\mathcal{H}_e(t)=& \frac{1}{2}\left(
  \begin{array}{cc}
     \Delta'& \Omega_{12}(t)\\
   \Omega_{12}^*(t)& -\Delta'\\
\end{array}\right),\\
\end{split}
\end{equation}
where $\Omega_{12}(t)=2\sqrt{2}g_{12}J_1(\beta)e^{-{\rm i}\chi(t)}$ and $J_1(\beta)$ is the first order Bessel function; therefore, the coupling strength can be adjusted through the parameter $\beta$. 
Clearly, Eq. (\ref{56}) exhibits the same structure as Eq. (\ref{hamiltonian}), suggesting that the state $|11\rangle$ can accrue a geometric phase via evolution, i.e., $|11\rangle\rightarrow e^{{\rm i}\gamma}|11\rangle$. Consequently, within the  two-qubit  subspace $\{|00\rangle, |01\rangle, |10\rangle,|11\rangle\}$, we can realize a nontrivial two-qubit geometric controlled-phase ($CP$) gate, i.e.,
\begin{equation}
\label{57}
U_{12}=|00\rangle\langle00|+|01\rangle\langle01|+|10\rangle\langle10|+e^{{\rm i}\gamma}|11\rangle\langle11|.
\end{equation}
Next, as a representative example, we optimize the performance of the $CP$ gate with $\gamma= \pi/2$. The parameter settings are $g_{12}=2\pi\times4.5$ MHz, $\Gamma^i_-=\Gamma^i_z=2\pi\times2$ kHz ($i=1,\ 2$), $\Delta_{12}=2\pi\times700$ MHz, $\alpha_1=2\pi\times300$ MHz, and $\alpha_2=2\pi\times200$ MHz. Accounting for decoherence, the Rabi error $g'_{12}=(1+\epsilon) g_{12}$, and detuning error $\eta g_{12}(|11\rangle\langle11|+|22\rangle\langle22|)/2$, we derive the optimal parameter settings to enhance the performance of the $CP$ gate; the obtained optimal parameters are listed in Case 5 of Table \ref{Table1}. With these parameters, the fidelity of the geometric $CP$ gate in ONGQG scheme  surpasses $99.5\%$ across almost the entire error range, far exceeding that of the single-loop NGQG scheme, as depicted in Figs. \ref{fig12}(e) and \ref{fig12}(f).

\section{Conclusions}

In summary, we present a general protocol for constructing geometric quantum gates with on-demand trajectories using a single smooth pulse. This general approach successfully overcomes the inherent limitations found in conventional NGQG schemes. Furthermore, our approach allows for further optimization under different scenarios, significantly enhancing the performance of the quantum gate, including its fidelity and robustness. Moreover, our scheme can readily extend to other quantum systems, such as trapped ions, nitrogen-vacancy centers in diamond, and Rydberg atoms. This adaptability reinforces the potential of our protocol as a promising avenue for realizing high-fidelity and strong-robust geometric quantum gates, offering valuable insights for the advancement of large-scale quantum computation.

\acknowledgements
This work was supported by the National Natural Science Foundation of China (Grant No. 12275090), the Guangdong Provincial Key Laboratory (Grant No. 2020B1212060066), and the Innovation Program for Quantum Science and Technology (Grant No. 2021ZD0302303).

\appendix
\section{The single-loop NGQG Scheme}
In this section, we detail the construction of nonadiabatic geometric gates using the conventional NGQG scheme \cite{loop4,loop6}. For a resonant driven two-level quantum system, in the interaction picture, the interacting Hamiltonian    is given by
\begin{equation}
\label{A1}
H_c(t)=\begin{pmatrix} 0 &  \Omega_c(t) e^{-{\rm i}\phi_c}\\
\Omega_c(t) e^{{\rm i}\phi_c} & 0 \end{pmatrix},
\end{equation}
where $\Omega_c(t)=\Omega_m\textrm{sin}^2(\pi t/ T)$ and $\phi_c$ denote the amplitude and phase of the driving field, respectively.
For the conventional NGQG scheme, 
the entire evolution time $T$ is divided into three segments to ensure a geometric evolution. At the intermediate times $T_1$ and $T_2$, the pulse area and relative phase $\phi_c$ satisfy
\begin{equation}
\label{ }
\begin{cases}
  \int_0^{T_1}\Omega_c(t)dt =\theta_c,  \quad \phi_c=\phi-\frac{\pi}{2}, \quad \quad  \  \  t\in[0,T_1], \\
  \\
  \int_{T_1}^{T_2}\Omega_c(t)dt =\pi,  \quad \phi_c=\phi+\gamma+\frac{\pi}{2}, \quad   t\in[T_1,T_2],  \\ 
  \\
  \int_{T_2}^{T}\Omega_c(t)dt =\pi-\theta_c,  \quad \phi_c=\phi-\frac{\pi}{2}, \quad  \  t\in[T_2,T].  \\
     \end{cases}
\end{equation}
Thus, the evolution operator at the final time is
\begin{eqnarray}
\label{E3}
U_c(T) &=& U_c(T,T_2)U_c(T_2,T_1)U_c(T_1,0) \notag\\
  &=&\cos\gamma+{\rm i}\sin\gamma\begin{pmatrix} \cos\theta_c & \sin\theta_c e^{-{\rm i}\phi}\\
\sin\theta_c e^{{\rm i}\phi} & -\cos\theta_c \end{pmatrix} \notag\\
&=&e^{{\rm i}\gamma \mathbf{n} \cdot \mathbf{\sigma} }
\end{eqnarray}
where $\mathbf{n}=(\sin\theta_c\cos\phi, \sin\theta_c\sin\phi, \cos\theta_c)$ is a unit directional vector, and $\mathbf{\sigma}=(\sigma_x, \sigma_y, \sigma_z)$ is a vector of Pauli operators. The evolution operator at the final time can also be denoted as
\begin{eqnarray}
\label{ }
U_c(T) &=&e^{{\rm i}\gamma}|\mu_+\rangle\langle \mu_+|+e^{-{\rm i}\gamma}|\mu_-\rangle\langle \mu_-|,
\end{eqnarray}
where $|\mu_+\rangle=\cos(\theta_c/2)|0\rangle +\exp{({\rm i}\phi)}\sin(\theta_c/2)|1\rangle$ and $|\mu_-\rangle=\exp{(-{\rm i}\phi)}\sin(\theta_c/2)|0\rangle-\cos(\theta_c/2)|1\rangle$ are the eigenstates of $\mathbf{n} \cdot \mathbf{\sigma}$.
Clearly, the operator $U_c(T)$ represents a rotational gate with the rotational axis $\mathbf{n}$ and rotational angle $\gamma$. Since the parallel-transport condition is satisfied, i.e., $\langle\mu_{\pm}|U_c^{\dag}(t)H_c(t)U_c(t)|\mu_{\pm}\rangle=0$, $U_c(T)$ is a geometric gate;
therefore, any single-qubit geometric gate can be realized by selecting appropriate parameters $\{\theta_c,\ \phi,\ \gamma\}$. For instance, the $T$ gate and $H$ gate correspond to the parameters $\{0,\ 0,\ \pi/8\}$ and $\{\pi/4,\ 0,\ \pi/2\}$, respectively.

%\nocite{*}%显示所有bib上的文献
%	\bibliography{ref.bib} %.bib文件名字
%	\bibliographystyle{ly} %.bst模板

%

\end{CJK}
\end{document}